\documentstyle[12pt,aasms4]{article}

\journalid{337}{15 January 1989}
\articleid{11}{14}
\def\simlt{\mathrel{\hbox to 0pt{\lower 3.5pt\hbox{$\mathchar"218$}\hss}
      \raise 1.5pt\hbox{$\mathchar"13C$}}}
  \def\simgt{\mathrel{\hbox to 0pt{\lower 3.5pt\hbox{$\mathchar"218$}\hss}
      \raise 1.5pt\hbox{$\mathchar"13E$}}}

\begin{document}

\title{Angular Sizes of Faint Field Disk Galaxies:\\
Intrinsic Luminosity Evolution}

\author{Laura Cay\'on, Joseph Silk}
\affil{Astronomy Department and Center for Particle
    Astrophysics, University of California,
    Berkeley, CA 94720}
\centerline \&
\author {St\'ephane Charlot}
\affil{Institut d'Astrophysique du CNRS, 98~bis boulevard Arago, 75014 Paris, France}

\begin{abstract}

In order to explain the small scale-lengths detected in the
recent deep field observations performed from large ground-based
telescopes and from the Hubble Space Telescope, we investigate
the predictions at high redshifts for disk galaxies that formed 
by infall. Changes with redshift in the observed properties of 
field galaxies are directly related to the evolution of the disks
and of the 
stellar populations.  We see that changes in the rest frame 
luminosity of a galaxy induce smaller values of half-light radii 
than are predicted assuming no evolution. Comparisons are presented
with two observed samples from Mutz et al.\markcite{mutz} (1994) 
and Smail et al.\markcite{smail} (1995).

\end{abstract}


\section{Introduction}

Studies of the faint field population of galaxies have traditionally
been based on number counts and analyses of redshift distributions.
Distinguishing Hubble types and/or measuring the morphological
characteristics of these galaxies has only become feasible with
the advent of new era telescopes. Observations from the largest 
ground-based telescope have provided data on the sizes of faint 
field galaxies (Smail et al.\markcite{smail}1995), while more 
refined morphological classification can be achieved using
high-resolution observations with the Hubble Space Telescope
(Mutz et al.\markcite{mutz} 1994, Casertano et al.\markcite{casertano} 
1995, Driver, Windhorst \& Griffiths\markcite{driver} 1995, Glazebrook 
et al.\markcite{galzeb} 1995, Driver et al.\markcite{driver2} 1995, 
Schade et al.\markcite{schade} 1995, Abraham et al.\markcite{abraham} 1996). 

At faint magnitudes $I\approx22$ and $R\approx26$ (expected median 
redshift greater than $0.5$; e.g. Lilly et al.\markcite{lilly} 1995), 
the dominant field population consists of very small disk systems with mean
scale-length of $\sim 0.2-0.3$ arcsec (Casertano et al.\markcite{casertano} 
1995, Smail et al.\markcite{smail} 1995). The corresponding drop in median 
galaxy size with increasing magnitude and redshift is faster than 
predicted by assuming a fixed intrinsic angular size of galaxies (Mutz 
et al.\markcite{mutz} 1994, Im et al.\markcite{im} 1995, Smail et 
al.\markcite{smail} 1995). Hence, faint distant galaxies are either 
intrinsically smaller or their emission is more concentrated toward 
the nuclear regions than that of low-redshift bright galaxies. In fact,
the evolution of stellar populations in galaxies is likely to affect 
the intrinsic surface brightness profile and therefore the observed 
half-light radius. 

Studies of our own galaxy  have converged on a model for global star formation  in which the Galaxy formed inside out, leading to predictions of star formation rate, gas surface density and metallicity as functions of time and galactic radius, {\it e.g.} Lacey and Fall\markcite{laf} (1985); Wang and Silk\markcite{was} (1994). Prantzos and Aubert\markcite{pa} (1995) have demonstrated that a robust and generic model contains 
the following ingredients: a rate of star formation that is proportional to the product of gas surface density and differential rotation rate,
and  infall of unenriched gas  during the first few Gyr. The 
radial dependence of the star formation prescription on total gas surface density,
 and via a star formation threshold on differential rotation rate, has been  verified for a sample of nearby spirals in a study of $H_\alpha$ emission from prominent HII regions by Kennicutt\markcite{ken1} 1989. 
Because 
the rotation curves are flat throughout most of the star-forming regions of disks, it follows that disks  must have formed inside-out. Hence 
small scale-lengths are expected at 
early ages.

In 
this {\it Letter}, we consider idealized disk galaxy models with 
infall to explore the influence of the radial dependence of the 
star formation rate on the observed angular sizes of distant galaxies. 
In \S2 we present models that reproduce the characteristics of local 
disk galaxies of various morphological types and compute the predicted
evolution of the intrinsic surface brightness profiles with redshift. 
In \S3 we compare our models with recent observations by Mutz et 
al.\markcite{mutz} (1994) and Smail et al.\markcite{smail} (1995). We
conclude from this simple but quantitative analysis that intrinsic 
luminosity evolution provides a potential explanation for the decrease
in the angular size of faint galaxies observed in deep surveys. 
In all the calculations presented below we adopt a fixed Hubble constant 
of $H_o=50$ km/sec/Mpc.

\section{Disk Galaxies: Infall Formation}

We first describe infall formation models that reproduce
the observed characteristics of present-day disk galaxies of
various Hubble types. Models with infall are favored by recent 
studies of chemical evolution in the Galactic disk that are 
consistent with the observed gas and star surface densities, metallicity
profile, and star formation rate (Ferrini et al.\markcite{ferrini} 1994;
Dopita \& Ryder\markcite{dopita} 1994; see Prantzos \& Aubert\markcite{pa} 
1995 for a detailed comparative analysis). Prantzos \& Aubert conclude 
that a phenomenological model satisfying the observational requirements is
one with infall in which the SFR depends on radius and time following a
Schmidt-type law
\begin{equation}
SFR(r,t)=(1-R)^{-1}{{\Sigma_g(r,t)}\over {\tau_g(r)}}~~~
{\rm M}_{\odot} {\rm pc}^{-2} {\rm Gyr}^{-1},
\end{equation}
\noindent
where $R\approx 0.32$ is the returned fraction of mass that was formed 
into stars, $\Sigma_g(r,t)$ is the gas surface density at radius $r$ and 
age $t$, and $\tau_g(r)=[0.3(1-R)(r_{\odot}/r)]^{-1}$~Gyr is the gas
consumption timescale ($r_{\odot }=8.5$ kpc). This model explains many properties of our galaxy both as a function of radius and of age.
Equation (1) relies on 
the instantaneous recycling approximation, which must be relaxed in 
studies of the metallicity in the inner radii of disks (e.g., Prantzos \& 
Aubert\markcite{pa} 1995). Since we are interested here only in galactic 
luminosity and size evolution, our results are not appreciably affected 
by this approximation.  The evolution of gas surface density in the infall 
disk model can be written
\begin{equation}
{{d\Sigma_g(r,t)}\over {dt}}=-SFR(r,t)(1-R)+f(r,t), 
\end{equation}
\noindent
where $f(r,t)\propto \exp[-t/\tau_f(r)]$ is the infall rate of metal-free
gas with characteristic timescale $\tau_f(r)$. We normalize the infall 
rate to the total surface density $\Sigma_{tot}$ of stars plus gas
observed at the age $T=13.5$ Gyr. Solving equation (2) for the evolution 
of the gas surface density, we obtain
\begin{equation}
\Sigma_g(r,t)=\Sigma_{tot}(r,T){{exp[-t/\tau_g(r)]-exp[-t/\tau_f(r)]}\over
 {\{1-[\tau_f(r)/\tau_g(r)]\}\{1-exp[-T/\tau_f(r)]\}}}.
\end{equation}

So far we have considered a pure disk model, for which the main
free parameters are 
the infall time scale, $\tau_f(r)$ and 
the total surface density distribution of
stars plus gas at time $T$, $\Sigma _{tot}(r,T),$.
We now adjust these parameters for galaxies
of different Hubble types (Sb, Sbc-Sc, and Sd-Im) by combining the
disk model with a type-dependent bulge component to reproduce the 
observed spectral energy distributions, surface brightness profiles, 
bulge-to-total luminosity ratios, and Scalo parameters (i.e., the 
ratio of present-to-past averaged SFR) of nearby galaxies. To achieve 
this, we  compute the spectrophotometric properties of model galaxies,
using the latest version of the Bruzual-Charlot stellar population 
synthesis models (Charlot, Worthey, \& Bressan\markcite{charlot} 1996).
The initial mass function is assumed to be a power law with the Salpeter 
slope and upper and lower cutoffs at $125\,M_\odot$ and $0.1\,M_\odot$,
respectively. We fix the metallicity at the solar value. This
should be a good approximation at most ages but may slightly
underestimate (by less than 20\%) the luminosity at the youngest
ages because metal-poor stars are somewhat more luminous than
metal-rich ones (Bruzual \& Charlot\markcite{bruzual} 1996).

We adopt constant infall time scales of 
$\tau_f=3$~Gyr, 7~Gyr, and $\infty$ 
for the model Sb, Sbc-Sc, and Sd-Im galaxies, respectively.  
For all galaxy types, at $T=13.5$~Gyr the disk component is assumed
to have the total surface density distribution of stars plus gas 
observed in the disk of the Galaxy (Fig.~1 of Prantzos \& 
Aubert\markcite{pa} 1995). We then define the surface
density of stars plus gas inside the bulge (at $T=13.5$ Gyr) 
as well as the 
bulge radius 
for the 
different types by requiring that (1) the bulge-to-total luminosity 
ratios of Sb, Sbc-Sc, and Sd-Im galaxies be 0.20, 0.13, and 0.0, 
respectively (Kennicutt et al.\markcite{kenn} 1994, Ferguson \& 
McGaugh\markcite{ferguson} 1995) and (2) the bulge and disk surface 
brightness profiles fit exponential laws with a type-independent 
scale length ratio of $\sim 0.1$ (Andredakis et al.\markcite{andredakis}
1995; Courteau et al.\markcite{courteau} 1996). 
The model galaxy bulges at $T=13.5$ Gyr are assumed to have
null gas surface density
and exponential stellar surface 
density profiles normalized to
fit the adopted disk profiles at bulge radii of 2 kpc and $1.5$ kpc
for the Sb and Sbc-Sc types, respectively.
The validity of the models is checked 
by comparing the
model spectra at age 13.5~Gyr with those of present-day Sb, 
Sbc-Sc, and Sd-Im galaxies. The spectral energy distribution is 
essentially controlled by the Scalo parameter, which is observed
to range from about 0.01 in Sa galaxies to over 1.0 in late-type 
spiral galaxies (e.g., Kennicutt et al.\markcite{kenn} 1994). 
The Scalo parameters produced by the adopted 
Sb, Sbc-Sc and Sd-Im models are $\sim 0.45$, $\sim 0.7$ and
$\sim 1.2$ respectively.  
The model Sb, Sbc-Sd, and Sd-Im galaxies 
computed in this way have absolute $B$ magnitudes at $T=13.5$~Gyr of 
about $-19.5$, $-19.7$, and $-19.9$, respectively.

The predicted rest frame surface brightness profiles at an age
of 13.5~Gyr are displayed in Figure~1a for the three galaxy types.
Figure~1b shows that the evolution of the underlying stellar 
population implies significant changes in the surface brightness
profiles. At $z=3$, the central regions of the Sb galaxy are 
intrinsically significantly brighter than at $z=0$, while those
of the the Sd-Im galaxy are only slightly brighter. The reason
for this is that early-type disk galaxies with short timescales
of star formation fade more rapidly than late-type ones. As a
result, from $z=0$ to $z=3$, the apparent surface brightness of
Sb, Sbc-Sc, and Sd-Im galaxies drops by about 3, 4, and 5~mag/arcsec$^2$,
respectively. As we shall see below, these changes in the surface
brightness profiles of disk galaxies with infall have important
consequences for the observed angular sizes and half-light radii.
In Figure~1c we have convolved the predicted profiles with an
observational point spread function (PSF) of FWHM 0.1~arcsec. We
note the consistency between our predicted profiles for forming
bulge/disk systems and the profiles fitted to star-forming galaxies
at redshifts greater than 3 (Giavalisco et al.\markcite{giav} 1996).

\section{Comparison with Observations}

Sizes of faint field galaxies have been measured in high-resolution 
HST images as part of the Medium Deep Survey (MDS) project
(Griffiths et al.\markcite{griff} 1994, Mutz et al.\markcite{mutz} 1994,
Casertano et al.\markcite{casertano} 1995, Phillips et al.\markcite{phillips}
1995, Forbes et al.\markcite{forbes} 1995), which allow morphological
classification to $I\simlt22$~mag, and, to even fainter magnitudes ($R<27$
mag), in ground-based studies using the Keck telescope (Smail et
al.\markcite{smail} 1995). Even though ground-based studies do not allow
morphological classification of galaxies at these faintest magnitudes, a
comparison with the models presented in \S2 is also important because disk 
galaxies are expected to dominate the field population (see below). In
comparing our models with observations, we account for the effect of the 
PSF and specific magnitude measurement scheme adopted (aperture or 
isophotal).  The convolution with a PSF has the effect of increasing the 
inferred half-light radii. We further assume for simplicity that the 
simulated galaxies are observed face-on. We now present results of 
comparisons of the models with the observations of Mutz et 
al.\markcite{mutz} (1994) and Smail et al.\markcite{smail} (1995).

Consider first data from the HST MDS project, taken with the Wide Field
Planetary Camera in two filters $V(5550~{\AA})$ and $I(8930~{\AA})$ (Mutz
et al.\markcite{mutz} 1994).  These include isophotal magnitudes measured
down to outer isophotes of $SB(V)\sim 27.0$ mag/arsec$^2$ and $SB(I)\sim
25.5$~mag/arcsec$^2$ for 33 morphologically identified disk galaxies at
$0.02\simlt z\simlt0.6$ (PSF of $FWHM\sim 0.1$ arcsec). In Figure 2 we
compare the observed evolution of half-light radius and $I$ magnitude 
as a function of redshift with the predictions of the various disk models.
We have assumed that galaxies formed at a redshift $z_f=10$ in an 
$\Omega=1$ universe, but similar results would be obtained for 
cosmologies with $z_f=5-30$ and $\Omega=0.1-1.0$. Also indicated by
large crosses in the two panels are the observed properties of nearby 
disk galaxies from the sample of Mathewson et al.\markcite{mathew}
(1992). The predicted angular sizes and $I$ magnitudes are consistent with
the mean of the observations at low and high redshifts, although the model $I$
magnitudes may be slightly fainter than the MDS magnitudes at the highest
redshifts. Figure~2a also shows that at $z\simgt1$, the predicted half-light 
radius of Sd-Im galaxies becomes smaller than that of earlier-type disks.
The reason for this is that Sd-Im galaxies have the faintest central surface 
brightness (Fig.~1b), and hence only light from the very central part of the
disk satisfies the isophotal magnitude selection. For comparison, we also
show in Figure~2a the predictions of no-evolution models assuming constant
intrinsic luminosity profiles of galaxies. Even though evolution affects
only mildly the predictions in the observed range of redshifts, it makes
considerable difference at $z\simgt1$.
 
Next, we consider the sample of galaxies observed in the $V$, $R$, and
$I$ bands by Smail et al.\markcite{smail} (1995) with the Keck Telescope.
Following Smail et al., we compute model $R$ magnitudes by adopting
the largest of a fixed 1.5~arcsec diameter aperture or isophotal diameter
down to a surface brightness limit of $SB(R)\approx28.3$ mag/arcsec$^2$.
The adopted seeing FWHM is 0.6~arcsec. The observed mean half-light radii
are plotted as a function of $R$ magnitude in Figure~3 together with model
predictions for four different combinations of $z_f$ and $\Omega$.
The observational relation is more tightly defined than the radius-redshift
and magnitude-redshift relations considered previously (Fig.~2) because 
it is based on a much larger sample of objects. As a result, the 
differences in the predictions for the various disk Hubble types are 
larger than the observational uncertainties.

To make a more meaningful comparison with the observations,
we have estimated weighted average values of the predicted half-light radius
of the disk population as a function $R$ magnitude. We assume that $70~\%$
of the local field population is accounted for by disk galaxies
with ratios Sb$:$Sbc-Sc$:$Sd-Im$\sim 25 \% :25 \% :20 \% $ (Ferguson \& 
McGaugh\markcite{ferguson} 1995) independent of redshift. We do not include
here the contribution to the average half-light radius values by elliptical
and S0 galaxies, but a recent morphological analysis of ultradeep HST
images suggests that early-type galaxies should account for only a
minority of the observed galaxies at the faintest magnitudes (Abraham et
al.~1996). Although a large fraction of galaxies at $I\simgt25$~mag in these
observations appear to be peculiar, irregular, or merging systems,
this may be in part a bias due to our viewing these systems in
rest-frame $U/B$ bands. We then expect the high-redshift predictions of
our models, presented in terms of half-light radii rather than detailed
profile fitting, to be a reasonable representation of disk sizes in young
or even forming galaxies. Figure~3 shows results both including and not
including intrinsic luminosity evolution. The differences in the shapes of
the $r_{\rm hl}$ curves are especially striking at the faintest $R$
magnitudes. The predictions of models with evolution to $R\approx26$ 
depend moderately on the assumed redshift of formation and density parameter.
The dependence is stronger for Sb galaxies than for Sd-Im galaxies because
earlier-type galaxies have shorter timescales of star formation and fade
more significantly as they age. As expected, predictions of models
without evolution do not fit the data at faint magnitudes, and they depend
more markedly on $\Omega$. We interpret the overall agreement between
predictions and observations in Figure~3 as an indication that, 
within the framework of disk galaxy formation by infall, 
intrinsic
luminosity evolution 
should be included
to explain, at least partially, the 
small observed sizes of disk galaxies at faint magnitudes.

In summary, we have applied infall formation models to simulate the 
radial dependence of the SFR in different Hubble types of disk galaxies.
The inside-out formation of galaxies in these models,
together with the chemical evolution of their stellar populations,
change 
the
rest-frame surface brightness profiles of the simulated galaxies as
they evolve. Earlier types are more affected by this intrinsic luminosity
evolution. Observational constraints for the samples analyzed in Mutz 
et al.\markcite{mutz} (1994) and in Smail et al.\markcite{smail} (1995)
were imposed on the modeled galaxies.  As expected and shown in Figures~2
and 3, {\em 
infall
models including rest frame luminosity evolution predict angular
sizes smaller than models with fixed half-light radius}. Moreover, such 
models can explain the clear trend toward very small sizes observed at 
faint magnitudes and high redshift. The simple models considered here 
are based primarily on fitting the properties of nearby galaxies and 
provide a robust description of present epoch star formation rates and 
luminosity profiles. Evolving these models backward in time is the most 
direct approach towards interpreting data on distant galaxies. It would 
be of considerable interest to refine the predictions of our models by 
including a more realistic mix of galaxy types and luminosities in order 
to make a more detailed comparison with the characteristics of the 
population observed in the Hubble Deep Field sample.

\acknowledgments
We acknowledge Ian Smail for providing us with the data presented in Figure~3. 
This work was supported in part by a grant from NASA.

\newpage	

\figcaption{Apparent surface brightness profiles as a function
or physical radius for Sb (solid line), Sbc-Sc (dotted line) and
Sd-Im (dashed line) model galaxies with infall at ({\it a}) $z=0$
and ({\it b}) $z=3$, for $\Omega=1$, $H_0=50$~km~s$^{-1}$Mpc$^{-1}$,
and a formation redshift $z_f=10$. ({\it c}) same as ({\it b}) but
as a function of angular radius and after convolving the model
profile with a PSF of $FWHM=0.1$ arcsec.}


\figcaption{Observed ({\it a}) half-light radius
and ({\it b}) $I$ magnitude as a function of redshift. The
circles are from the observed sample of faint disk galaxies 
by Mutz et al. (1994), while large crosses indicate the range
of properties of nearby disk galaxies (Mathewson et al. 1992).
The curves show the predictions for the Sb ({\it solid}), Sbc-Sc
({\it long dashed}), and Sd-Im ({\it short dashed}) models
both with ({\it thick}) and without ({\it thin}) intrinsic
luminosity evolution. Cosmological parameters are the same as 
in Fig.~1.}


\figcaption{Observed half-light radius versus apparent $R$ magnitude.
The data are from the observations of Smail et al. (1995), filled
circles corresponding to medians of 401 galaxies, horizontal
error bars to the extent of the magnitude bin, and vertical error bars 
to the $95\%$ confidence limits (see original paper for detail). Model
predictions are shown for four different cosmologies, ({\it a}) 
$z_f=30, \Omega =1.0$, ({\it b}) $z_f=5, \Omega =1.0$, ({\it c}) 
$z_f=30, \Omega =0.1$, 
and ({\it d}) and $z_f=5, \Omega =0.1$. Solid (thin) lines denote
the predicted half-light radii versus R magnitude for the Sb model. 
Long dashed lines refer to the Sbc-Sc model and short dashed lines 
correspond to the Sd-Im model predictions.  Evolution of the intrinsic
luminosity has been considered in these predictions. The thick solid 
lines correspond to the weighted average of the half-light radius 
predicted in the different cosmologies for the disk field population. 
Thin dashed-dotted lines show corresponding weighted averages for 
the models with fixed half-light radius (no intrinsic luminosity
evolution).}


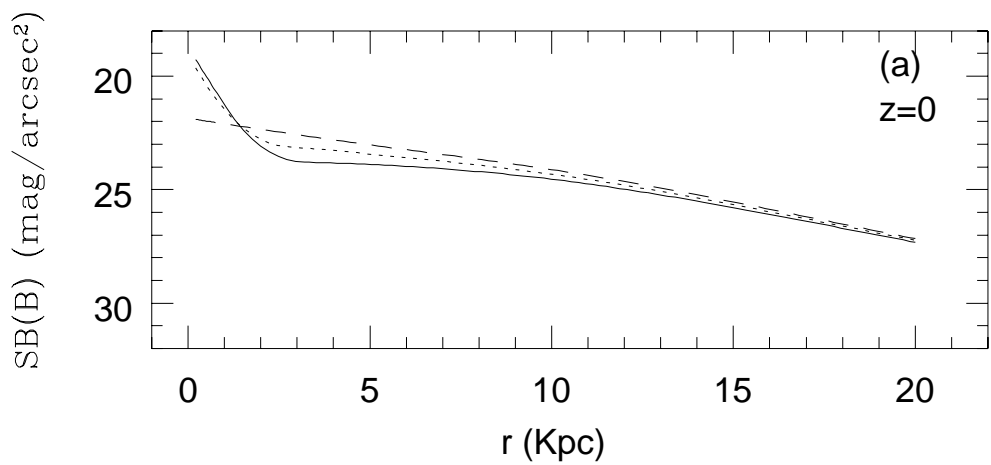
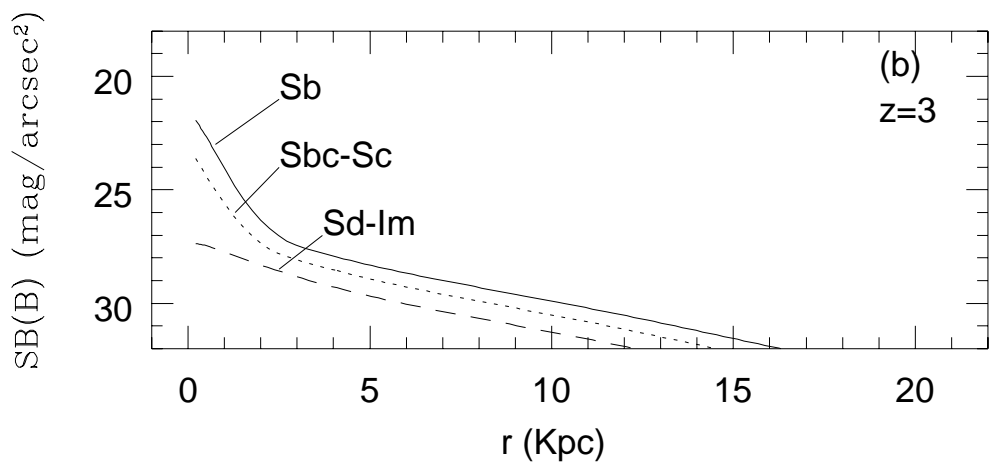
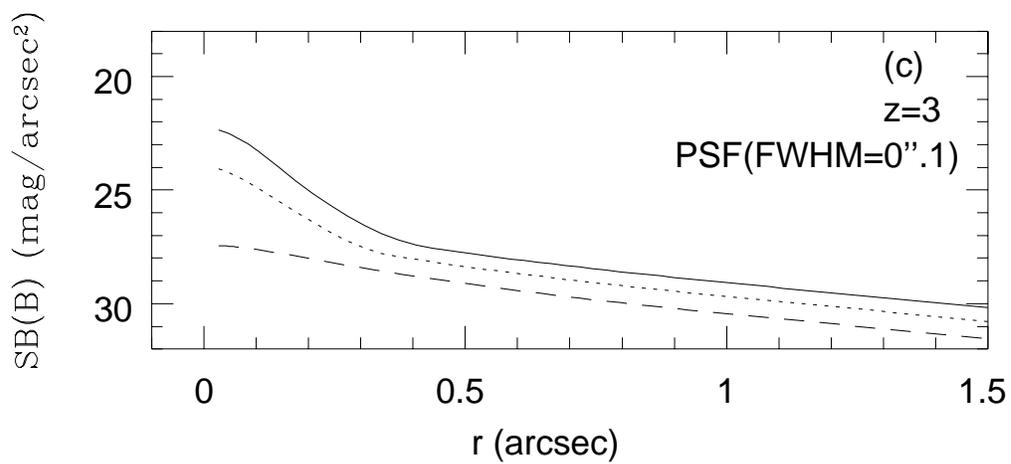
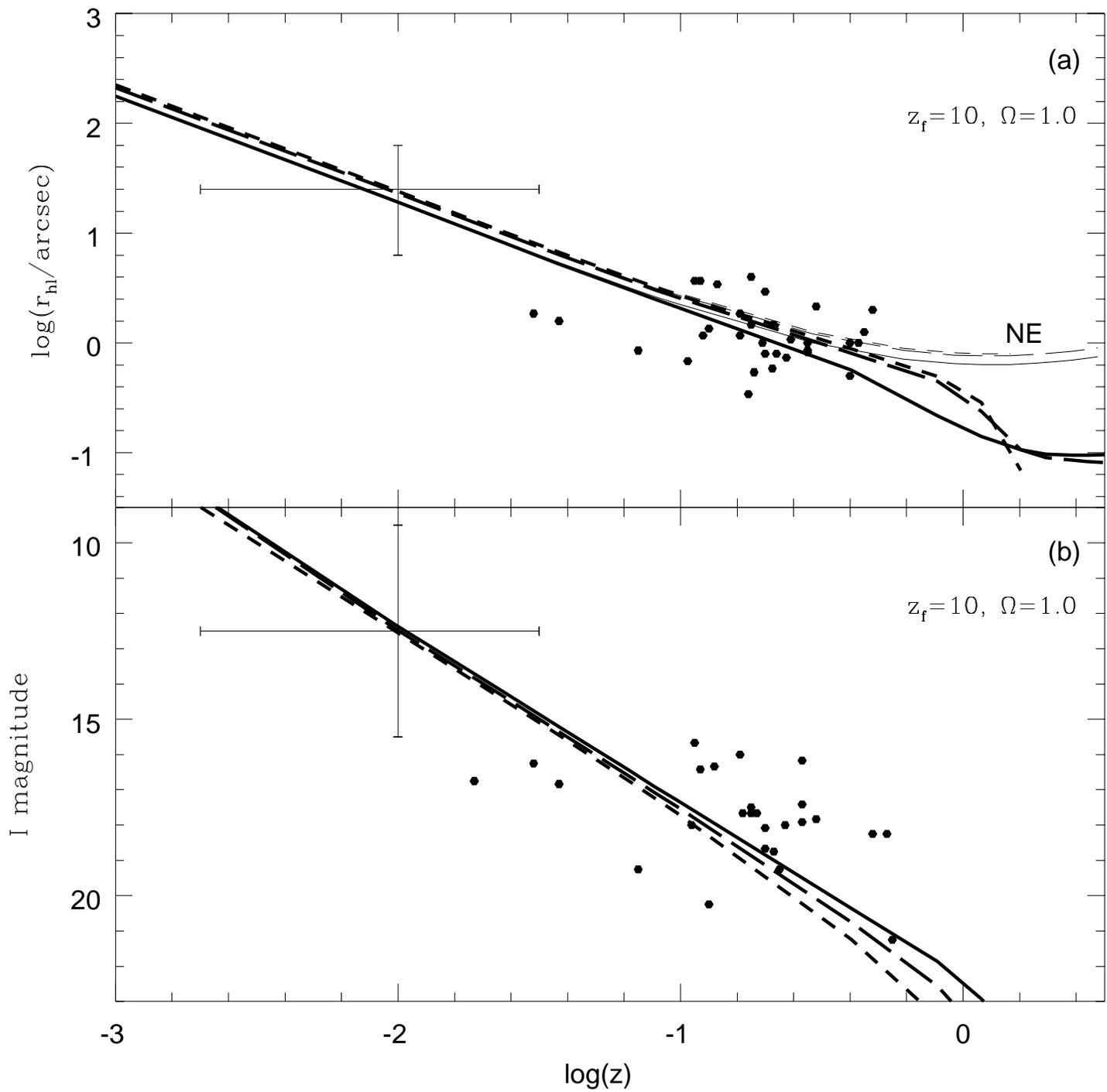
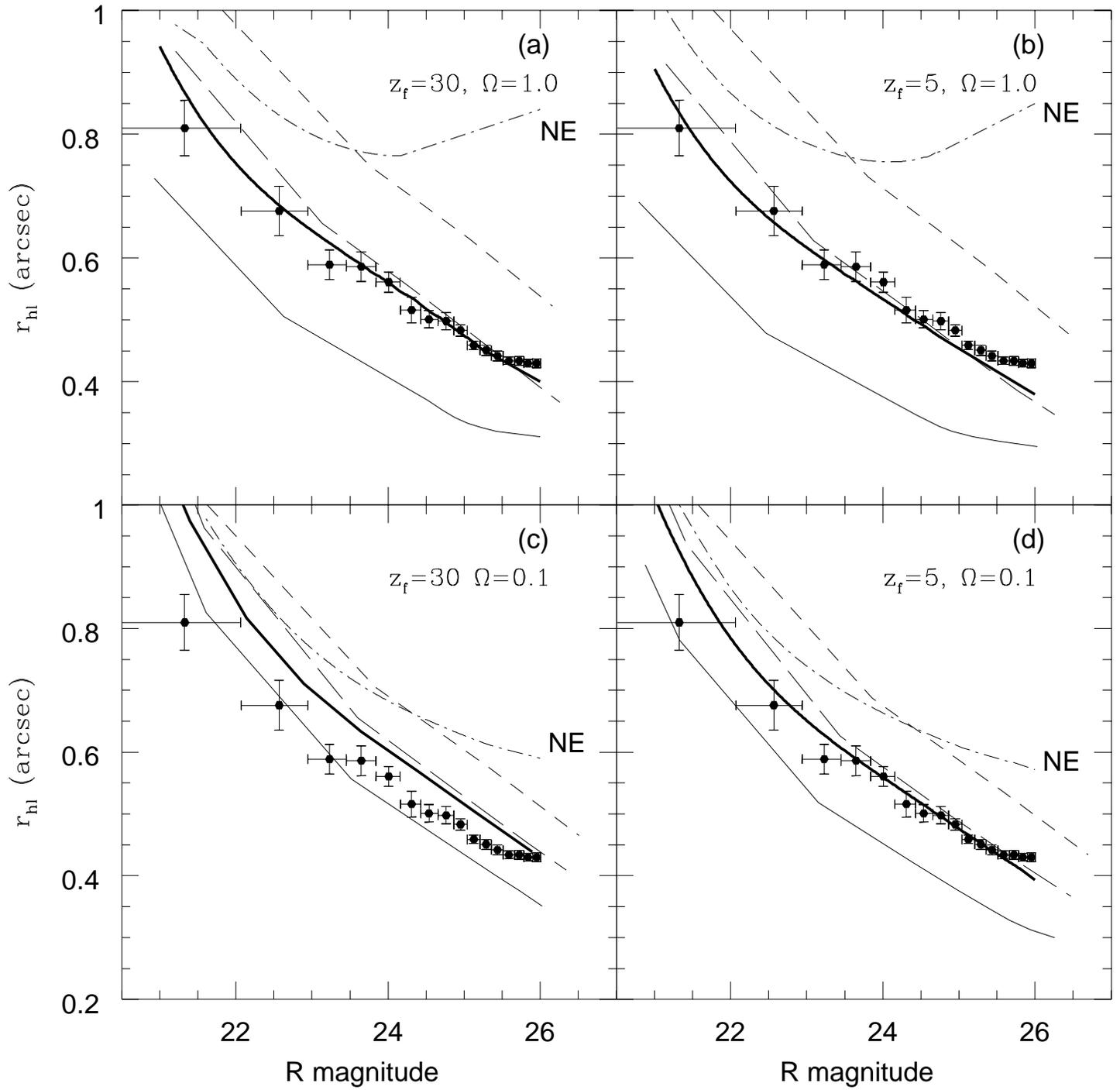
\end{document}